\documentclass[a4paper,11pt]{article}
  
\usepackage{amsmath,amssymb}
\usepackage{graphicx}
\usepackage{natbib}
\usepackage{booktabs}%
\usepackage{url}
\usepackage[top=1in,bottom=1in,left=1in,right=1in]{geometry}

\newcommand{\norm}{\mathrm{N}}
\newcommand{\gam}{\mathrm{Gam}}
\newcommand{\bet}{\mathrm{Beta}}
\newcommand{\bern}{\mathrm{Bern}}

\newcommand{\vect}{\mathrm{vec}}
\def\T{{ \mathrm{\scriptscriptstyle T} }}

\renewcommand{\vec}[1]{\boldsymbol{#1}}
\newcommand{\vecn}[1]{\boldsymbol{#1}}
\newcommand{\matr}[1]{\mathnormal{#1}}

\usepackage{tikz}

\DeclareRobustCommand\dashed{\tikz[baseline=-0.6ex]\draw[thick,dashed] (0,0)--(0.54,0);}

\newcommand{\myPhi}{\mathsf{\Phi}}
\newcommand{\myZ}{\mathsf{Z}}
\newcommand{\edge}{\relbar\mkern-7mu\relbar\mkern-7mu\relbar}
\newcommand{\diredge}{\longrightarrow}
\usepackage{arydshln}
\newcommand{\VAR}[2]{\textsc{VAR}$_{#1}\text{(}{#2}\text{)}$} 
\newcommand{\VARG}[2]{\textsc{VAR}$\text{(}{#1}\text{, }{#2}\text{)}$} 
\newcommand{\ie}{\emph{i.e.}\ }
\newcommand{\eg}{\emph{e.g.}\ }

\usetikzlibrary{arrows.meta}

\tikzset{
    regulations/.cd,
    act/.style={-{Stealth}}, % Style for activation
    rep/.style={-{Bar}}, % Style for repression
} 

\newcommand{\regulationarrow}[1][act]{%
    \tikz[baseline]{\draw[regulations/#1] (0,0.5ex) --++ (1.5em,0);}%
}

\definecolor{ggplot5_1}{HTML}{F8766D}
\definecolor{ggplot5_2}{HTML}{A3A500}
\definecolor{ggplot5_3}{HTML}{00BF7D}
\definecolor{ggplot5_4}{HTML}{00B0F6}
\definecolor{ggplot5_5}{HTML}{E76BF3}

\usepackage{enumerate}
\usepackage{caption}
\usepackage{subcaption}

\makeatletter
\renewcommand\p@subfigure{\thefigure~}
\makeatother

\providecommand{\keywords}[1]
{
  \small
  \textbf{\textit{Keywords---}} #1
}

\title{Bayesian inference of sparsity in stable vector autoregressive processes}
\author{Sarah E. Heaps$^1$\footnote{\texttt{sarah.e.heaps@durham.ac.uk}}, Ian H. Jermyn$^1$, Yujiang Wang$^2$ and Darren J. Wilkinson$^1$}
\date{$^1$Department of Mathematical Sciences, Durham University, U.K.\\
$^2$School of Computing, Newcastle University, U.K.}

\begin{document}

\maketitle

\begin{abstract}
Advances in sensing technology have made it possible to collect large volumes of high-dimensional time-series data. In fields like genetics and neuroscience, key questions concern whether directed relationships between variables can be learned from these data. To this end, graphical vector autoregressions are a popular tool because zeros among the autoregressive coefficients and error precision matrix have natural interpretations in terms of Granger non-causality and contemporaneous conditional independence. In applications where system dynamics are subject to functional or structural constraints, assuming the process is stable can be advantageous. However, enforcing stability demands restricting the autoregressive coefficients to lie in a constrained space with a complex geometry called the stationary region. The resulting inferential challenges are compounded when sparsity is also a requirement. Working in the Bayesian paradigm, we tackle the problem of developing a prior that simultaneously enforces stationarity and sparsity through parameter expansion, constructing a spike-and-slab prior with support constrained to the stationary region. A mixture of $G$-Wishart distributions provides a sparse prior for the error precision matrix. Computational inference is carried out using Metropolis-within-Gibbs, exploiting the No-U-Turn Sampler and reversible-jump steps. We demonstrate the inferential and predictive benefits of our approach through simulations and applications in macroeconomics and neuroscience.
\end{abstract}


\section{Introduction}\label{sec:intro}
Over recent years, advances in sensing technology have made it possible to collect large volumes of time-series data on many variables. In a diverse array of fields, such as systems biology \citep[][]{MO20}, genetics \citep[][]{MA13}, neuroscience \citep[][]{GFOBC13} and finance \citep[][]{ABC16}, there is substantial interest in how these data can be used to improve understanding of the interactions between variables. %For example, time-series comprising multielectrode recordings of brain activity can be used to characterise the connections between regions of interest in the brain. This facilitates contributions to a wider discussion of the role of such connectivity patterns in mental health disorders and neurological conditions. 
Exploiting the time-ordering of data, \citet{Gra69} proposed the framework of Granger causality in order to endow these interactions with a notion of causality: informally, a process $y_{1}$ is Granger causal for another process $y_{2}$ if utilising information on the history of $y_{1}$ improves the prediction of $y_{2}$. In this paper, we focus on Bayesian inference for the class of stable graphical vector autoregressions, which have an interpretation in terms of strong Granger causality \citep[][]{Eic12}. This is formulated through conditional independence statements, which can be represented as a mixed graph $G$. The graphical representation is appealing because it facilitates effective communication of a potentially large number of inter-variable relationships; see \citet{SF22} for a Granger causality review.

Vector autoregressions are a popular class of models for multivariate time-series data. An $m$-variate, zero-mean Gaussian process $\vec{y}_t$ evolving according to a vector autoregression of order $p$, or \VAR{m}{p} process, can be written as
\begin{equation}\label{eq:var}
\vec{y}_{t} = \matr{\phi}_1 \vec{y}_{t-1} + \ldots + \matr{\phi}_p \vec{y}_{t-p} + \vec{\epsilon}_t,
\end{equation}
where the errors $\vec{\epsilon}_t$ form an uncorrelated sequence of zero-mean Gaussian random vectors, $\vec{\epsilon}_t | \matr{\Sigma} \sim \norm_m(\vecn{0}, \matr{\Sigma})$. The parameters of the model therefore comprise the autoregressive coefficient matrices $\matr{\phi}_s \in M_{m \times m}(\mathbb{R})$ ($s=1,\ldots,p$) and the error variance matrix $\matr{\Sigma} \in \mathcal{S}^+_m$, where $M_{m \times n}(V)$ denotes the set of $m \times n$ matrices with entries in $V$ and $\mathcal{S}^+_m$ denotes the set of $m \times m$ symmetric, positive definite matrices; henceforth, we refer to the collection of $\matr{\phi}_s$ as $\myPhi$. The graphical interpretation arises by allowing sparsity in both the autoregressive coefficient matrices $\myPhi$ and the inverse of the error variance matrix, $\matr{K}=\matr{\Sigma}^{-1}$. Here, non-zero entries in the autoregressive structure correspond to directed edges in the graph and represent statements of Granger causality, while non-zero entries in the error precision matrix correspond to undirected edges, representing statements of contemporaneous conditional dependence. In the Bayesian paradigm, zeros can be accommodated through a sparse prior distribution, typically borrowing ideas from the Bayesian variable selection literature for the (unconstrained) coefficients $\myPhi \in M_{m \times m}(\mathbb{R})^p$ and from the Gaussian graphical modelling literature for the positive definite error precision $\matr{K} \in \mathcal{S}^+_m$.% For example, sparsity in the elements of the autoregressive coefficient matrices is often facilitated through spike-and-slab priors \citep[][]{GSN08} or continuous shrinkage priors coupled with a criterion for truncation to zero \citep[][]{HHNCAGW23}. 

Over the last decade, the modelling framework for graphical vector autoregressions has been extended to accommodate multiple realisations of multivariate time-series in hierarchical or panel vector autoregressions \citep[][]{CGYHSV17,Kor16}. To reduce the computational burden of model fitting, there have also been extensions in a variety of directions, such as variational Bayes approaches as an alternative to Markov chain Monte Carlo (MCMC) \citep[][]{BBB24}. However, much of the focus has been on exploiting structural constraints in graphical vector autoregressions to achieve dimension reduction, for example by structuring the autoregressive coefficient matrices into a three-way tensor and then applying a tensor decomposition of reduced rank \citep[][]{FSCS22}. For an overview of recent advancements in the broader field of vector autoregressions, we refer to \citet{Luo25}. 

Using the backshift operator $B$, defined such that $B \vec{y}_t = \vec{y}_{t-1}$,~\eqref{eq:var} can be written as $\vec{\epsilon}_t = (\matr{I}_m - \matr{\phi}_1 B - \cdots - \matr{\phi}_p B^p) \vec{y}_t = \matr{\phi}(B) \vec{y}_t$ where $\matr{I}_m$ is the $m \times m$ identity matrix and $\matr{\phi}(u) = \matr{I}_m - \matr{\phi}_1 u - \cdots - \matr{\phi}_p u^p$, $u \in \mathbb{C}$, is termed the characteristic polynomial. A vector autoregression is stable if and only if all the roots of $\det \{ \matr{\phi}(u) \} = 0$ lie outside the unit circle. We refer to this subset of $M_{m \times m}(\mathbb{R})^p$ as the stationary region, denoted $\mathcal{C}_{p,m}$. From a practical perspective, an important property of stationary processes is that the variances of forecasts do not grow without bound as the forecast horizon increases. Particularly in the life sciences, where system dynamics can be subject to functional or structural constraints, this is often a desirable property. This includes applications where interest lies in long-term forecasting or when studying a linear dynamic system which is thought to be at equilibrium. Stationarity can also be advantageous for the visualisation of results: for a broad class of graphical time series model that are piecewise stationary, \citet{Eic12} introduced various Granger-causal Markov properties that enhance graphical interpretation by relating separation properties on the graph to statements of conditional independence or Granger non-causality. Motivated by these arguments, stationarity is frequently stated as an assumption in the literature on graphical vector autoregressions. However, it is generally only enforced as a constraint over a simpler subset of $\mathcal{C}_{p,m}$ \citep[][]{HSW13} or, as highlighted in \citet{DYMZ23}, not enforced as a constraint at all. Yet, because the number of parameters in $\myPhi$ grows quadratically with dimension $m$, for a time-series of fixed length, there is typically increasing epistemic uncertainty in the values of the $\phi_{s,ij}$, as $m$ or the order $p$ get large, allowing posterior mass to spill outside the stationary region. The problem is particularly pronounced if the length of the time-series is short relative to $m$, which makes it difficult to infer, with near certainty, that the process is stationary \citep[][]{Hea23}. 

In principle, the solution to this problem is simple: develop a prior distribution for the autoregressive coefficient matrices which can simultaneously induce sparsity while constraining inference to the (full) stationary region. In practice, however, it is difficult to specify a prior with these properties which facilitates tractable computational inference due to the geometric complexity of the stationary region, compounded by the complications arising from allowing sparsity in the parameter space. The main contribution of this paper is to address precisely this challenge. Introducing an auxiliary parameter, we define a bijective mapping between the parameter-expanded set of stationary autoregressive coefficient matrices and a parameter-expanded set of unconstrained square matrices. This mapping is constructed in such a way that zeros in the unconstrained square matrices map to zeros in the autoregressive coefficient matrices, and vice versa. As a result, specification of a sparsity inducing prior for the unconstrained square matrices produces a sparsity inducing prior for the autoregressive coefficient matrices that is constrained to the stationary region. To allow sparsity in the error precision matrix, we adopt a prior comprising a mixture of $G$-Wishart distributions \citep[][]{Rov02,AM05} which, to the best of our knowledge, has not been applied previously in the context of graphical vector autoregressions. Crucially, we show that computational inference for this reparameterised model requires only minor modifications to the algorithm that would be used for an analogous model that ignores the stationarity constraint. We describe an efficient Metropolis-within-Gibbs algorithm to sample from the posterior, which uses the No-U-Turn sampler \citep[][]{HG14} to draw from the full conditional distribution of the continuous parameters pertaining to the reparameterised autoregressive coefficient matrices, and the
%direct double conditional Bayes factor sampler \citep[][]{HLHG14} 
graphical model determination algorithm \citep[][]{WL12} to draw from the full conditional distribution of the error precision matrix. Through an extensive simulation experiment, we highlight the benefits of enforcing stationarity, and suggest situations in which the benefit is likely to be most pronounced. We then consider two substantive applications using data from macroeconomics and neuroscience, which demonstrate the improved predictive performance and scientific insight that can be gained by fitting our stationary graphical vector autoregression.

The remainder of this paper is organised as follows. In Section~\ref{sec:var}, we introduce graphical vector autoregressions, then motivate and develop our sparsity-inducing, stationarity-enforcing prior for the autoregressive coefficient matrices. We also describe the mixture of $G$-Wishart priors for the error precision matrix. Section~\ref{sec:mcmc} outlines our MCMC scheme for computational inference. In Section~\ref{sec:sim_expt}, we detail an extensive simulation experiment before considering applications from macroeconomics and neuroscience in Section~\ref{sec:applications}.

\section{\label{sec:var}Graphical vector autoregressions}

\subsection{The graphical model}
Let $G = (V, E_1, E_2)$ denote a mixed graph where $V = \{1, \ldots, m \}$ denotes the set of vertices, $E_1  \subseteq V \times V$ is a set of directed edges and $E_2  \subseteq \binom{V}{2}$ is a set of undirected edges. We denote a directed edge from $a \in V$ to $b \in V$ by $a \diredge b$ and an undirected edge between $a \in V$ and $b \in V$ by $a \edge b$. Two distinct vertices, $a$ and $b$, can be unconnected or connected by any combination of a directed edge in either direction and an undirected edge. A stationary graphical \VAR{m}{p} process~\eqref{eq:var} associated with the graph $G$, denoted by \VARG{p}{G}, has parameters that are constrained by:
\begin{enumerate}[(i)]
\item $a \diredge b \notin E_1$ if and only if $\phi_{s,ba}=0$ for all $s=1,\ldots,p$ in which case we say that $y_a$ is (strongly) Granger non-causal for $y_b$;
\item $a \edge b \notin E_2$ if and only if $K_{ab}=K_{ba}=0$ in which case we say that $y_a$ and $y_b$ are contemporaneously conditionally independent;
\end{enumerate}
for all $a \ne b$. It follows that non-zero entries in off-diagonal elements of $\matr{\phi}_{s}$, for any $s=1,\ldots,p$, lead to directed edges, reflecting Granger causality relationships, while non-zero entries in the off-diagonal elements of $\matr{K}$ correspond to undirected edges, reflecting contemporaneous conditional dependence structure. %\citet{Eic12} showed that the graph $G$ satisfies various Granger-causal Markov properties that can be applied to relate separation properties on the graph to statements of contemporaneous conditional independence or Granger non-causality. %\citet{PC20} illustrate the types of insights that this can yield by using a simpler model, which only allows sparsity in $\matr{K}$, to learn an undirected spatial network. 

\subsection{\label{subsec:prior_Phi}Prior for autoregressive coefficient matrices}

\subsubsection{Parameter expansion on an unconstrained space}
As our inferential objective is to characterise Granger causality structure, we consider priors for the autoregressive coefficient matrices $\myPhi$ which encourage zeros only in the off-diagonal elements $\phi_{s,ij}$ for $i \ne j$, $s=1,\ldots,p$. When stationarity is not enforced, a typical approach would introduce a vector of unknown hyperparameters $\vec{\vartheta}^\T=(\vec{\vartheta}_1^\T, \ldots, \vec{\vartheta}_p^\T)$ and adopt a prior of the form
\begin{equation}
\pi(\myPhi, \vec{\vartheta}) = \prod_{s=1}^p \pi(\vec{\vartheta}_s) \prod_{i=1}^m \prod_{j=1}^m \pi(\phi_{s,ij} | \vec{\vartheta}_s),\label{eq:unconstrained_prior}
\end{equation}
where $\pi(\phi_{s,ij} | \vec{\vartheta}_s)$ for $i \ne j$ is a Bayesian variable selection prior and the role of $\vec{\vartheta}_s$ is to allow the data to influence the degree of selection, shrinkage, or both, at lag $s$. For example, $\pi(\phi_{s,ij} | \vec{\vartheta}_s)$ might be a spike-and-slab prior in which the slab is a normal distribution, centred at zero, and the spike is either another zero-mean normal distribution with smaller variance \citep[][]{GSN08} or an atom of probability at zero \citep[][]{LBGGW11}. In these cases, the unknown hyperparameters $\vec{\vartheta}_s$ would typically include the precision of the slab and the mixing probability. Alternatively,  $\pi(\phi_{s,ij} | \vec{\vartheta}_s)$ might be a continuous shrinkage prior, such as the horseshoe, in which case $\vec{\vartheta}_s$ would typically include local and global shrinkage parameters \citep[][]{HHNCAGW23}. A naive approach to modifying this prior to constrain inference to the stationary region would simply truncate the density~\eqref{eq:unconstrained_prior} such that $\myPhi \in \mathcal{C}_{p,m}$. However, this would introduce an intractable normalising constant that depended on $\vec{\vartheta}$, making the posterior distribution doubly-intractable and hence significantly complicating computational inference. In addition, even if $\vec{\vartheta}$ was fixed, the geometric complexity of the stationary region $\mathcal{C}_{p,m}$ would make it difficult to design an efficient proposal distribution for $\myPhi$ that restricted most of its mass to $\mathcal{C}_{p,m}$.

Rather than truncating a prior that is designed for unconstrained Euclidean space, we present a more tailored solution. The basic idea is first to find an unconstrained reparameterisation of a stationary process and then to apply a sparsity-inducing prior over the reparameterised space. \citet{Hea23} uses an unconstrained reparameterisation of a stationary vector autoregression to construct a prior which restricts inference to the stationary region. However the new parameters are partial autocorrelation matrices where zeros do not have a graphical interpretation in terms of Granger non-causality. In order to construct a reparameterisation in which the meaning of individual zeros is preserved, we adopt an approach based on parameter-expansion \citep[][]{LW99,MV99}. Although originally introduced as an auxiliary variable technique to accelerate Gibbs sampling algorithms, parameter-expansion has been used more recently to aid specification of priors over constrained regions such as the Stiefel manifold \citep[][]{JHD21} or spaces of rank-reduced positive semi-definite matrices \citep[][]{HJ24}.

Introducing the auxiliary variable $\nu \in \mathbb{R}^+$, we map $(\myPhi, \nu) \in \mathcal{C}_{p,m} \times \mathbb{R}^+$ to $(\myZ, u) \in M_{m \times m}(\mathbb{R})^p \times (0, 1)$, where $\myZ = (\matr{Z}_1, \ldots, \matr{Z}_p)$, through
%\begin{subequations} \label{eq:generalisedVAR}
%\begin{align}
%\matr{Z}_s & = \nu^s \matr{\phi}_s \quad \text{for $s=1,\ldots,p$} \\
%\intertext{and} 
%u & = \rho(\matr{C}_{\phi})
%\end{align}
%\end{subequations}
\begin{equation}\label{eq:generalisedVAR}
\matr{Z}_s  = \nu^s \matr{\phi}_s, \ \text{$s=1,\ldots,p$;} \quad  u  = \rho(\matr{C}_{\phi}),
\end{equation}
where $\rho(\cdot)$ denotes the spectral radius and $\matr{C}_{\phi}=\matr{C}(\matr{\phi}_1,\ldots,\matr{\phi}_p)$ is the companion matrix
\begin{equation*}
\matr{C}_{\phi} =
\begin{pmatrix}
\matr{\phi}_1 &\matr{\phi}_2 &\cdots &\matr{\phi}_{p-1} &\matr{\phi}_p\\
\matr{I}_m    &\matr{0}      &\cdots &\matr{0}          &\matr{0}\\
\matr{0}      &\matr{I}_m    &\cdots &\matr{0}          &\matr{0}\\
\vdots        &\vdots        &\ddots &\ddots            &\ddots\\
\matr{0}      &\matr{0}      &\cdots &\matr{I}_m        &\matr{0}\\
\end{pmatrix}.
\end{equation*}
It can easily be shown (\eg by induction) that the reciprocals, $\eta_\phi$, of the non-zero eigenvalues of the companion matrix $\matr{C}_{\phi}$ are the roots of $\det \{ \matr{\phi}(\eta_\phi) \} = 0$ and so the stability condition is satisfied if and only if $\rho(\matr{C}_{\phi}) < 1$. If we introduce a corresponding companion form for the unconstrained matrices, $\matr{C}_{z}=\matr{C}(\matr{Z}_1,\ldots,\matr{Z}_p)$, it can similarly be shown that the reciprocals, $\eta_z$, of the non-zero eigenvalues of the companion matrix $\matr{C}_z$ satisfy
\begin{equation*}
0 = \det\left(\matr{I}_m - \matr{Z}_1 \eta_z - \ldots - \matr{Z}_p \eta_z^p\right) = \det\left\{\matr{I}_m - \matr{\phi}_1 (\nu \eta_z) - \ldots - \matr{\phi}_p (\nu \eta_z)^p\right\}.
\end{equation*}
If follows that all (non-zero) eigenvalues of $\matr{C}_z$ are of the form $\nu \lambda_\phi$ where $\lambda_\phi = 1 / \eta_\phi$ is an eigenvalue of $\matr{C}_\phi$. Therefore, because $\nu > 0$, $\rho(\matr{C}_z) = \nu \rho(\matr{C}_\phi)$ and we can interpret $\nu$ as a scaling factor which maps the Schur stable block matrix $\matr{C}_\phi$ to the block matrix $\matr{C}_z$ with unconstrained first row block. We can therefore write the inverse of~\eqref{eq:generalisedVAR} as
%\begin{subequations}\label{eq:generalisedVARinv}
%\begin{align}
%\matr{\phi}_s & = \frac{u^s \matr{Z}_s}{\rho(\matr{C}_z)^s} \\
%\intertext{for $s = 1,\ldots,p$, and} 
%\nu & = \frac{\rho(\matr{C}_z)}{u}.
%\end{align}
%\end{subequations}
\begin{equation}\label{eq:generalisedVARinv}
\matr{\phi}_s = \frac{u^s \matr{Z}_s}{\rho(\matr{C}_z)^s}, \ s = 1,\ldots,p; \quad \nu  = \frac{\rho(\matr{C}_z)}{u}.
\end{equation}
We note that the inverse mapping~\eqref{eq:generalisedVARinv} is not defined when $\rho(\matr{C}_z)=0$, that is, when $\matr{C}_z$ is a nilpotent matrix. As we discuss in Section~\ref{subsubsec:prior_zu}, we will adopt a prior in which the diagonal elements of the $\matr{Z}_s$ are continuous and so $\rho(\matr{C}_z)>0$ almost surely, making the mapping~\eqref{eq:generalisedVAR} almost surely bijective on the support of $(\myZ, u)$.

\subsubsection{Prior over parameter-expanded reparameterisation}\label{subsubsec:prior_zu}

For building a prior distribution for the autoregressive coefficient matrices, the reparameterisation described in the previous section has three important theoretical properties. First, for any off-diagonal element at any lag $s=1,\ldots,p$, $z_{s,ij}=0$ if and only if $\phi_{s,ij}=0$ ($i \ne j$). Therefore by assigning a joint prior to $(\myZ, u)$ which allows sparsity amongst the unconstrained $z_{s,ij}$ ($i \ne j$), we can induce a distribution for $(\myPhi, \nu)$ such that the marginal prior for the $\phi_{s,ij}$ has precisely the intended property of sparsity. Second, the Jacobian determinant of the transformation~\eqref{eq:generalisedVAR} can be calculated in closed form as $J\{(\myPhi, \nu); (\myZ, u)\} = - \nu^{q-1} \rho(\matr{C}_\phi)$, where $q=m^2 p (p+1)/2$; a proof can be found in Section~S1 
%\ref{smsec:jacobian} 
of the Supplementary Materials. This facilitates derivation of properties of the marginal prior induced for $\myPhi$, which can be used to guide the choice of prior for $u$, as we explain in Supplementary Section~S2. 
%\ref{smsec:constraintsonpriorparameters} 
Finally, results from simulations strongly suggest that shrinkage increases with lag, which is consistent with the commonly-held belief that the importance of lagged variables decays with lag length. In Supplementary Section~S3, 
%\ref{smsec:shrinkageatdifferentlags}  
we present some theoretical indications that support these observations.
% \footnote{I would like to make the claim that if the $\matr{Z}_s$ are given a prior with zero mean and a variance that is common across $s$, then the $\matr{\phi}_s$ with higher lag will be shrunk more rigorously towards zero. This may not always be true but it certainly seems to be in all the simulations I've looked at. It won't be possible to do an exact variance calculation and I don't think we can apply the delta method because of the different behaviour of the Jacobian matrix (not just determinant) in the complex eigenvalue versus real eigenvalue cases. Is there anything theoretical we can say? If not, I guess I could just say something about what simulations suggest.}

To each off-diagonal element in the set of unconstrained square matrices $z_{s,ij}$, $i \ne j$, $s=1,\ldots,p$, we assign a spike-and-slab prior, which is formulated by introducing an indicator variable $\gamma_{s,ij}$ and a continuous \emph{effect size} parameter $\tilde{z}_{s,ij}$ and by writing $\matr{Z}_s = \matr{\Gamma}_s \circ \matr{\tilde{Z}}_s$ where $\circ$ denotes the Hadamard product and the diagonal elements of the $\matr{\Gamma}_s$ are fixed at 1. Introducing hyperparameters $\vec{\vartheta} = (\tau, \vartheta, \omega)^\T$, we then adopt the prior
\begin{equation*}
\pi(u, \matr{\tilde{Z}}_1,\ldots,\matr{\tilde{Z}}_p, \matr{\Gamma}_1,\ldots,\matr{\Gamma}_p, \vec{\vartheta}) = \pi(u) \pi(\tau) \pi(\vartheta) \pi(\omega) \prod_{s=1}^p \pi(\matr{\tilde{Z}}_s | \tau, \omega) \pi(\matr{\Gamma}_s | \vartheta) ,
\end{equation*}
where $u \sim \bet(a_1, a_2)$ and, for $s=1,\ldots,p$,
\begin{alignat*}{3}
\tilde{z}_{s,ij} | \tau   &\sim \norm(0, \tau^{-1}),&     \quad \tau &\sim \gam(b_1, b_2),& \quad &i \ne j,\\
\gamma_{s,ij} | \vartheta &\sim \bern(\vartheta),&   \quad \vartheta &\sim \bet(c_1, c_2),& \quad &i \ne j,\\
\tilde{z}_{s,ii} | \omega &\sim \norm(\mu, \omega^{-1}),& \quad \omega &\sim \gam(e_1, e_2),& \quad &i=1,\ldots,m,
\end{alignat*}
where we will typically take $\mu=0$ to encourage shrinkage of the diagonal effect sizes to zero. However, an alternative is to assign $\mu \in \mathbb{R}$ a prior to allow more flexible shrinkage to a common, but unknown, value.

A computational advantage of formulating the spike-and-slab prior using a product of effect-size parameters and indicators is that it fixes the dimension of the continuous part of the parameter space, allowing all the effect-size parameters to be updated jointly using an efficient proposal distribution that exploits the gradient of their full conditional distribution; see Section~\ref{sec:mcmc} for further details. %Notational dependence on the matrices of indicators, $\matr{\Gamma}_1, \ldots, \matr{\Gamma}_p$, also allows straightforward construction of the adjacency matrix $\matr{W}_1=(w_{1,ij})$ encoding the subgraph, $G_1$, comprising all vertices but only the directed edges, $E_1$. Specifically, define $\matr{\tilde{W}}=(\tilde{w}_{ij})=\sum_{s=1}^p(\matr{\Gamma}_s-\matr{I}_m)^\T$, then the entries of $\matr{W}_1$ are given by $w_{1,ij}=\mathbb{I}(\tilde{w}_{ij} > 0)$, where $\mathbb{I}(x)$ is the indicator function, equal to 1 if $x$ is true and 0 otherwise.

\subsection{\label{subsec:prior_K}Prior for error precision matrix}

There have been various efforts to provide a sparse prior for the error precision matrix $\matr{K}$ in graphical vector autoregressions. Some approaches rely on a hyper-inverse Wishart prior for the error variance matrix $\matr{\Sigma}$ given its associated undirected graph. This distribution has an analytically tractable normalising constant only if the graph is decomposable, and so either inference has been limited to the space of decomposable graphs \citep[][]{PC20} or the calculations underpinning structural comparison have been approximated when graphs are not decomposable \citep[][]{CV05,MC09}. Other approaches map the error precision or stationary precision matrix to unconstrained Euclidean space through a Cholesky factorisation then assign sparse priors over the unconstrained space \citep[][]{GSN08,RRG24}. Although this greatly simplifies computational inference, the natural graphical interpretation of the pattern of sparsity is lost.

In order to avoid these limitations, we employ ideas from Gaussian graphical modelling, in which the problem of Bayesian inference for the sparsity structure of a precision matrix has been studied extensively; see \citet{Mas20} or \citet{VMS24} for recent reviews. In this literature, the $G$-Wishart distribution is often used as a conditional prior for the precision matrix, given the undirected graph, because it is conjugate to the Gaussian graphical likelihood. Specifying a marginal distribution over undirected graphs completes the prior by quantifying uncertainty in the sparsity pattern. Although the normalising constant of the $G$-Wishart density is also analytically intractable unless the graph is decomposable, its evaluation can be avoided during MCMC using ideas based on the exchange algorithm \citep[][]{MGM06}; see Section~\ref{sec:mcmc}. 

To this end, denote by $G_2$ the subgraph associated with all vertices but only the undirected edges $E_2$. We then specify a joint prior over $(G_2, \matr{K})$ as $\pi(G_2, \matr{K}) = \pi(G_2) \pi(\matr{K} | G_2)$ in which $\matr{K} | G_2 \sim \mathrm{W}_{G_2}(d, \matr{D})$. Here $\mathrm{W}_{G_2}(d, \matr{D})$ denotes the $G$-Wishart distribution with shape parameter $d > 2$, inverse scale matrix $\matr{D}$ and density $\pi(\matr{K} | G_2) = |K|^{(d-2)/2} \exp\left\{ -\frac{1}{2} \mathrm{tr}(\matr{K} \matr{D}) \right\} / I_{G_2}(d, \matr{D})$, in which $I_{G_2}(d, \matr{D})$ is the normalising constant. In the applications in this paper, we use a uniform distribution over the space of undirected graphs as the marginal prior for $G_2$ or, equivalently, independent uniform $\mathrm{U}\{0, 1\}$ distributions for the presence of each undirected edge. However, our approach could easily be modified to accommodate other options.

%\textit{Now say something about the G-Wishart distribution being a natural conjugate prior for a sparse precision matrix in a multivariate normal likelihood. However, until recently, its use has been hampered by the intractable normalising constant in its density which depends on the undirected graph. Lenkoski's direct sampler allows use of an exchange-type algorithm for computational inference which avoids the need to calculate it. These innovations do not seem to have reached the graphical VAR literature where, to the best of our knowledge, they have not been used, except in the special case of decompsable graphs, as detailed above. We buck the trend by employing the G-Wishart distribution as our prior for $\matr{K}$.} 

\section{\label{sec:mcmc}Posterior inference through MCMC}

Consider observations modelled as realizations from a stationary graphical vector autoregression of order $p$. The likelihood function associated with the observations at times $t=t_1,\ldots,n$ conditional on the preceding $p$ observations is given by
\begin{equation}\label{eq:likelihood}
p(\vec{y}_{t_1:n} | \vec{y}_{(t_1-p):(t_1-1)}, \myPhi, \matr{K}) = \prod_{t=t_1}^n p(\vec{y}_t | \vec{y}_{(t-p):(t-1)}, \myPhi, \matr{K})
\end{equation}
in which $\vec{y}_t \mid \vec{y}_{(t-p):(t-1)}, \myPhi, \matr{K} \sim \norm_{m}\left(\sum_{s=1}^p \matr{\phi}_s \vec{y}_{t-s} \, , \, \matr{K}^{-1}\right)$. In principle, the joint distribution of any $n$ consecutive observations is a zero-mean Gaussian distribution with analytically tractable variance matrix, $\matr{V}_0$. Therefore we could take $t_1=p+1$ and extend~\eqref{eq:likelihood} to construct the joint distribution of observations $\vec{y}_{1:n}$. Unfortunately, the variance matrix $\matr{V}_0$ for $\vec{y}_{1:p}$ would depend on the error precision matrix $\matr{K}$ in a manner that destroyed the semi-conjugacy of the $G$-Wishart prior. Modern algorithms for efficiently exploring the joint posterior for $(\matr{K}, G_2)$ exploit analytic simplifications that rely on the full conditional for $\matr{K}$ being a $G$-Wishart distribution. As such, we do not model the first $p$ observations through the stationary distribution. The simplest alternative is to simply take $t_1=p+1$ and condition on the first $p$ observations or, roughly equivalently, to model the first $p$ observations using a uniform distribution. This is the approach we take in the applications in this paper. However, if a time-series is particularly short, one might not want to lose completely the information contained in the first $p$ observations. In this case, one could set $t_1=1$, augment the parameter space with $n_1 \ge p$ (latent) initial values, $\vec{y}_{-(n_1-1)}, \ldots, \vec{y}_0$, and assign a prior $\vec{y}_t \sim \norm(\vec{0}, \matr{I}_m)$ for $t=-(n_1-1),\ldots,p-n_1$ while modelling $\vec{y}_{p-n_1+1},\ldots,\vec{y}_0$ through~\eqref{eq:likelihood}. For brevity in the remainder of this section, we assume that we have set $t_1=p+1$ and use the conditional likelihood $p(\vec{y}_{(p+1):n} | \vec{y}_{1:p}, \myPhi, \matr{K})$.

\begin{sloppypar}
We treat the likelihood as a function of the error precision matrix, $\matr{K}$, and the parameters that determine the autoregressive coefficient matrices, $(\matr{\tilde{Z}}_1,\ldots,\matr{\tilde{Z}}_p,\matr{\Gamma}_1,\ldots,\matr{\Gamma}_p,u)$. Combining it with the prior described in Sections~\ref{subsec:prior_Phi}-\ref{subsec:prior_K} through Bayes theorem yields the posterior distribution, $\pi(\matr{\tilde{Z}}_1,\ldots,\matr{\tilde{Z}}_p,\matr{\Gamma}_1,\ldots,\matr{\Gamma}_p, u, \vec{\vartheta}, \matr{K}, G_2 | \vec{y}_{1:n})$, which is proportional to
\end{sloppypar}
\begin{equation*}%\label{eq:posterior}
\pi(G_2) \pi(\matr{K} | G_2) \pi(\vec{\vartheta}) \pi(u) \left \{ \prod_{s=1}^p \pi(\matr{\tilde{Z}}_s | \vec{\vartheta}) \pi(\matr{\Gamma}_s | \vec{\vartheta}) \right\} p(\vec{y}_{(p+1):n} | \vec{y}_{1:p}, \myZ, u, \matr{K})
\end{equation*}
where $\matr{Z}_s = \matr{\Gamma}_s \circ \matr{\tilde{Z}}_s$, $s=1,\ldots,p$, and we recall that $\vec{\vartheta} = (\tau, \vartheta, \omega)^\T$. We note that the expanded parameter set $(\myZ, u)$ is not identifiable in the likelihood; in this case the likelihood depends only on $\myPhi$ but we can take any $\alpha > 0$ and replace $\matr{Z}_s$ with $\matr{Z}_s' = \alpha^s \matr{Z}_s \in M_{m \times m}(\mathbb{R})$ for $s=1,\ldots,p$ then clearly $\matr{\phi}_s' = (\alpha u)^s \matr{Z}_s / \{ \alpha \rho(\matr{C}_z) \}^s = \matr{\phi}_s$ so that the $\matr{\phi}_s$ are unchanged. Moreover, when any $\gamma_{s,ij}$ is equal to zero, the corresponding $\tilde{z}_{s,ij}$ cannot be determined from $z_{s,ij}$. Nevertheless, likelihood identifiability is not required to generate samples from the joint posterior of $(\matr{\tilde{Z}}_1,\ldots,\matr{\tilde{Z}}_p,\matr{\Gamma}_1,\ldots,\matr{\Gamma}_p,u)$ and transforming these samples through $\matr{Z}_s = \matr{\Gamma}_s \circ \matr{\tilde{Z}}_s$ and~\eqref{eq:generalisedVARinv} yields draws from the posterior of the identifiable parameters, $\myPhi$.

For the parameters pertaining to the coefficient matrices, we denote the vector of continuous parameters by 
%\begin{equation*}
$\vec{\theta}_{1,c} = \left\{ \vect(\matr{\tilde{Z}}_1)^\T, \ldots, \vect(\matr{\tilde{Z}}_p)^\T, \mathrm{logit}(u), \mathrm{logit}(\vartheta), \log(\tau), \log(\omega) \right\}^\T \in \mathbb{R}^{pm^2+4}$
%\end{equation*}
and the vector of unknown indicators by 
%\begin{equation*}
$\vec{\theta}_{1,d} = \left\{ \mathrm{offdiag}(\matr{\Gamma}_1)^\T, \ldots, \mathrm{offdiag}(\matr{\Gamma}_p)^\T \right\}^\T \in \{0, 1\}^{pm(m-1)}$.
%\end{equation*}
Here $\vect(\cdot)$ denotes the vec-operator, $\mathrm{logit}(x)=\log\{x / (1-x)\}$ and $\mathrm{offdiag}(\matr{X})$ is a column vector comprising the off-diagonal elements of a matrix $\matr{X}$, stacked by column. Finally, we denote the set of parameters arising from the undirected part of the graphical model by $\vec{\theta}_2 = (G_2, \matr{K})$. If we had augmented the parameter space with $n_1 \ge p$ (latent) initial values, $\vec{y}_{-(n_1-1)}, \ldots, \vec{y}_0$, these could simply be included in $\vec{\theta}_{1,c}$. In order to generate samples from the posterior distribution $\pi(\vec{\theta}_{1,c},\vec{\theta}_{1,d},\vec{\theta}_{2}| \vec{y}_{1:n})$, we adopt a Metropolis-within-Gibbs scheme, which repeatedly iterates through the following steps:
\begin{enumerate}
\item Sample $\vec{\theta}_{1,c}$ from its full conditional distribution (FCD) $\pi(\vec{\theta}_{1,c} | \vec{\theta}_{1,d}, \vec{\theta}_2, \vec{y}_{1:n})$ in a single block using the No-U-Turn sampler (NUTS) \citep[][]{HG14};
\item Sample the elements of $\vec{\theta}_{1,d}$ from their (Bernoulli) FCDs $\pi(\gamma_{s,ij} | \{\vec{\theta}_{1,d} \setminus \gamma_{s,ij}\}, \vec{\theta}_{1,c}, \vec{\theta}_2, \vec{y}_{1:n})$ in a systematic scan through each variable in turn;
\item Sample $\vec{\theta}_2$ from its FCD $\pi(\vec{\theta}_2 | \vec{\theta}_{1,c}, \vec{\theta}_{1,d}, \vec{y}_{1:n})$ using the graphical model determination algorithm of \citet{WL12}.
\end{enumerate}

The algorithm is implemented using JAX \citep[][]{JAX18}, which is a Python library providing an efficient tensor and array computation framework with support for automatic differentiation. The NUTS algorithm used in step~1 to sample from the full conditional density $\pi(\vec{\theta}_{1,c} | \vec{\theta}_{1,d}, \vec{\theta}_2, \vec{y}_{1:n})$ is an efficient variant of Hamiltonian Monte Carlo \citep[][]{Nea11}, which adaptively sets the path length tuning parameter. A gradient-based algorithm is particularly helpful here because of the complex posterior dependence amongst the $\matr{Z}_s$ that is induced by the mapping~\eqref{eq:generalisedVARinv}. The NUTS algorithm is implemented using BlackJAX \citep[][]{BLACKJAX24}. The step size and inverse mass matrix are determined using Stan’s Window Adaptation method \citep[][]{CGH17} based on a warm-up run in which all the unknown indicators in $\vec{\theta}_{1,d}$ are fixed at 1 and $\matr{K}$ is fixed at a (dense) estimate of $\matr{\Sigma}^{-1}$.

Explicit calculation of the FCDs for the $\gamma_{s,ij}$ and full details of the graphical model determination algorithm are provided in Supplementary Section~S4. In brief, the latter exploits the partial analytic structure of the $G$-Wishart distribution to simplify implementation of moves that attempt to adjust the graph, $G_2$. It also uses an exchange-type algorithm \citep[][]{MGM06,Lia10} to avoid calculation of the $G$-Wishart normalising constant which, in turn, requires a method for sampling from the $G$-Wishart distribution. This is achieved in a manner that avoids computationally expensive clique decomposition by employing an edgewise block Gibbs sampler. 
%We note that we also considered an algorithm that purports to generate direct samples from the $G$-Wishart distribution \citep[][]{Len13}. However, recent work suggests the algorithm is only approximate \citep[][]{TK25} $\ldots$ \footnote{I'm not sure whether I want to say anything further or, indeed, whether I want to mention this at all. I don't really want to present another set of results and I definitely don't want to wade into any debate about the correctness or otherwise of the Lenkoski algorithm. Shall I just delete the last two sentences?}

\section{\label{sec:sim_expt}Simulation experiment}

\subsection{Simulation set-up}
Denote by $\matr{W}_2=(w_{2,ij})$ the symmetric adjacency matrix encoding the undirected graph $G_2$ and by $\vec{w}_2=(w_{2,12}, w_{2,13}, w_{2,23}, \ldots, w_{2,m-1,m-1})^\T \in \{ 0,1 \}^{m(m-1)/2}$ its above diagonal elements. We are interested in the behaviour of (i) the marginal posterior distributions for the indicators $\vec{\theta}_{1,d}$ and $\vec{w}_2$ defining the sparsity patterns in the model parameters and (ii) the joint posterior distribution for the mixed graph $G$. To this end, we consider an idealised, hypothetical scenario in which data are simulated under a stationary graphical vector autoregression. In a series of experiments, we consider a range of values of $n \in \{200, 1000\}$ and $m \in \{5, 10, 20\}$ for the length and dimension of the time-series, respectively, and a range of values for the order $p \in \{1, 2, 4\}$ of the process. We also consider various proportions $r \in \{ 0.1, 0.5, 0.9 \}$ of sparsity in the model parameters, $\matr{\phi}_s$ and $\matr{K}$. For every combination of $n$, $m$, $p$ and $r$, we simulate ten sets of parameters and, from each, a $m$-variate time-series of length $n$, leading to 540 datasets in total. To simulate the autoregressive coefficient matrices, we generate independent samples: $u \sim \mathrm{U}(0, 1)$, $\gamma_{s,ij} \sim \bern(1-r)$ and $\tilde{z}_{s,ij} \sim \norm(0, 1)$ for $s=1,\ldots,p$, $i \ne j$, and $\tilde{z}_{s,ii} \sim \norm(0, 1)$ for $s=1,\ldots,p$, $i=1,\ldots,m$, and then construct the $\matr{\phi}_s$ through~\eqref{eq:generalisedVARinv}. To simulate the precision matrix, we generate independent samples $w_{2,ij} \sim \bern(1-r)$ for $i < j$, which determines $G_2$ and then draw $\matr{K} | G_2 \sim \mathrm{W}_{G_2}(3, \matr{I}_m)$.

For all analyses, we complete our prior from Sections~\ref{subsec:prior_Phi} and~\ref{subsec:prior_K} with the hyperparameter choices: $a_1=1$, $a_2=1$, $b_1=e_1=2.01$, $b_2=e_2=1.01$, $c_1=c_2=2$, $d=3$ and $\matr{D}=\matr{I}_m$. 
%In order to assess sensitivity to the choice of prior for $u=\rho(\matr{C}_\phi)$, we also considered inferences under the prior where $a_1=q+1$, $a_2=2$. We found $\ldots$

In order to investigate the effect on the posterior of enforcing stationarity through the prior, we compare our inferences to those obtained under the same model but with a sparsity-inducing prior that does not restrict inference to the stationary region. In this case, the $\matr{\phi}_s$ are assigned spike-and-slab priors of the same form as those chosen for the $\matr{Z}_s$ in Section~\ref{subsec:prior_Phi}, taking $\matr{\phi}_s = \matr{\Gamma}_s \circ \matr{\tilde{\phi}}_s$. The MCMC scheme is analogous to that in Section~\ref{sec:mcmc}, except the continuous model parameters simply comprise $\vec{\theta}_{1,c}=\left\{ \vect(\matr{\tilde{\phi}}_1)^\T, \ldots, \vect(\matr{\tilde{\phi}}_p)^\T, \mathrm{logit}(\vartheta), \log(\tau), \log(\omega) \right\}^\T \in \mathbb{R}^{pm^2+3}$. Computationally, this highlights the relative simplicity of our methodology for guaranteeing stationarity, requiring the addition of one auxiliary variable and the cost of moving between unconstrained and constrained parameter spaces. As a benchmark for the two Bayesian analyses, we compare against sparsity pattern estimates based on a regularised likelihood method, with lasso penalties on the off-diagonal elements in $\matr{\phi}_s$ and $\matr{K}$, selecting tuning parameters through a grid search using the extended Bayesian information criterion \citep[][]{CC08}. This is implemented by the \texttt{graphicalVAR} R package \citep[][]{Eps24}. We note that for twelve of the 540 datasets, numerical maximisation of the regularised log-likelihood function failed.

For all data sets, we use the MCMC algorithm in Section~\ref{sec:mcmc} to generate 25K samples from the posterior, after a burn-in of 25K draws, and thinning to every tenth iterate to reduce post-processing overheads. To evaluate the marginal posterior distributions of the indicators across experiments, we use the naive Bayes classifier to produce a point estimate of each indicator for each experiment, \ie\ we classify an indicator as 1 if its posterior probability exceeds 0.5 and 0 otherwise. For the $pm(m-1)$ indicators in $\vec{\theta}_{1,d}$ and the $m(m-1)/2$ indicators in $\vec{w}_2$, separately, the point estimates are compared against the values used to simulate the data. This yields a misclassification rate and the empirical distributions of these errors are then summarised across the ten experiments for each simulation setting. Of course, one of the main advantages of the Bayesian approach to inference is that we can fully quantify our uncertainty in each of the indicators through its posterior probability. Therefore we additionally consider these metrics for the subset of indicators in which we are more certain in our conclusions, only classifying an indicator as 1 or 0 if its posterior probability exceeds 0.9 or is less than 0.1, respectively. On average, 56\% of the indicators in $\vec{\theta}_{1,d}$ and 55\% in $\vec{w}_2$ were used in these latter comparisons.

\subsection{Results}
\subsubsection{\label{subsec:sim_expt_marginal}Marginal posteriors for indicators $\vec{\theta}_{1,d}$ and $\vec{w}_2$}
\begin{figure}[t!]
\centering
\begin{subfigure}{\textwidth}
\centering
\includegraphics[width=0.75\textwidth]{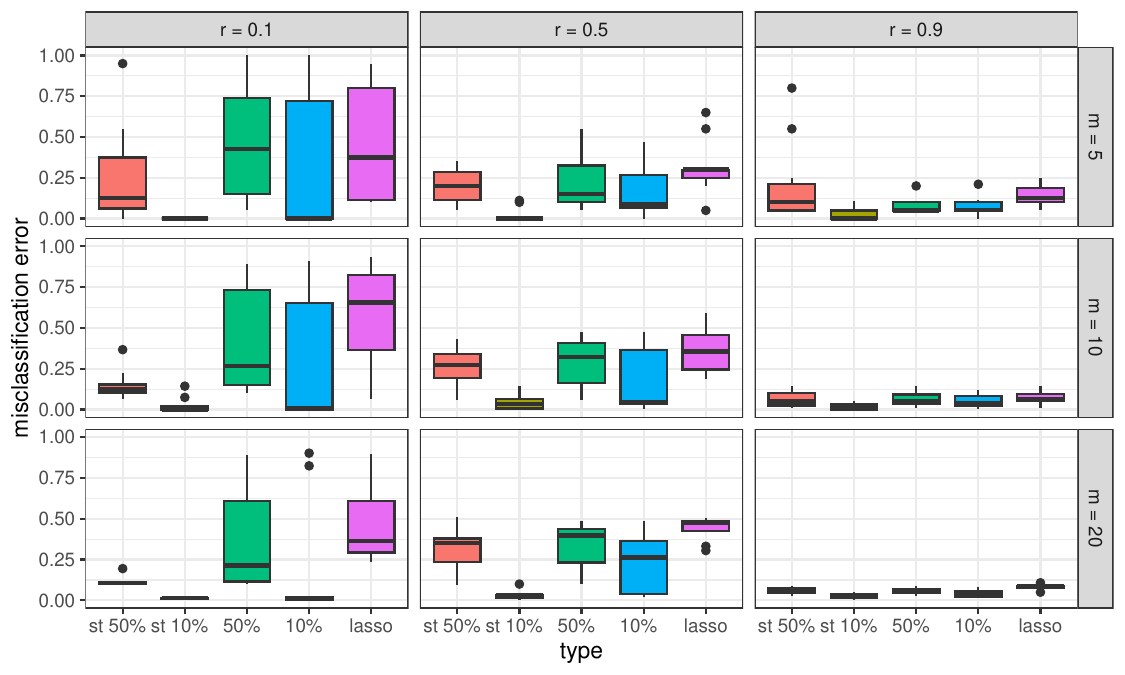}
\caption{}
\label{fig:boxplot_phi_n200_p1}
\end{subfigure}\\
\begin{subfigure}{\textwidth}
\centering
\includegraphics[width=0.75\textwidth]{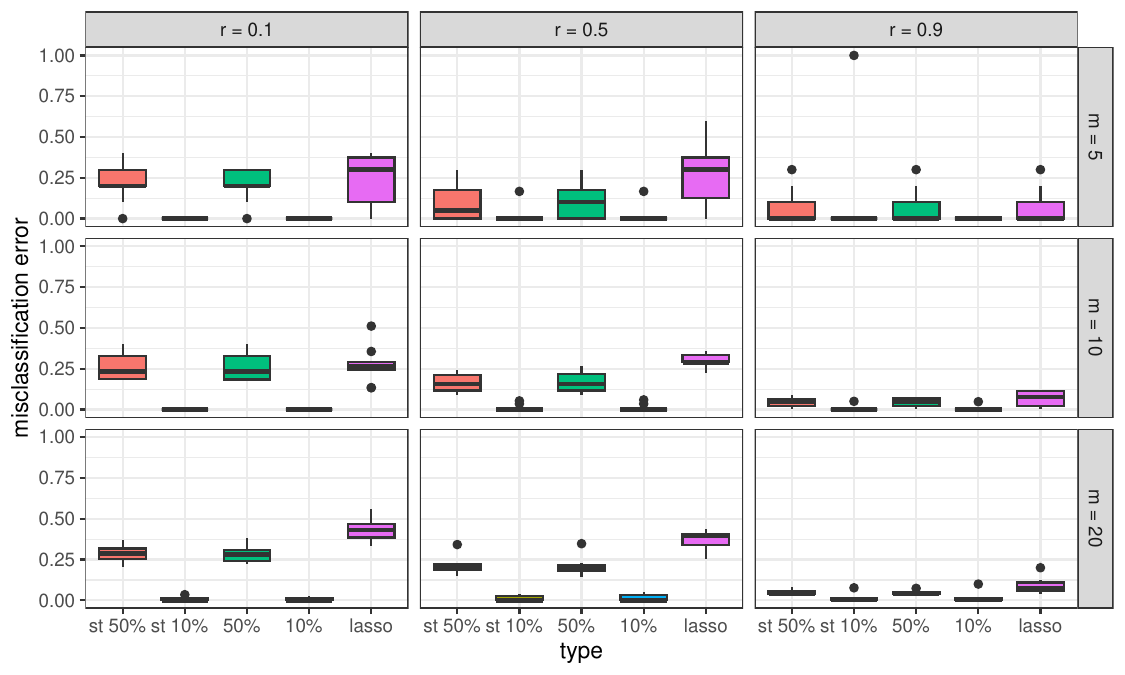}
\caption{}
\label{fig:boxplot_K_n200_p1}
\end{subfigure}
\caption{\label{fig:boxplots_n200_p1}Boxplots summarising misclassification rates for the indicators \subref{fig:boxplot_phi_n200_p1} $\vec{\theta}_{1,d}$ and \subref{fig:boxplot_K_n200_p1} $\vec{w}_2$ across ten simulated datasets when $n=200$, $p=1$ for various $m$ and $r$. Results are shown for the Bayesian analyses using naive Bayes classifiers when stationarity is (st 50\%) and is not (50\%) enforced and similarly when only evaluating classifications of indicators with posterior probability $>0.9$ or $<0.1$, (st 10\% and 10\%). Results are also shown for classifications based on the Lasso-based estimator (lasso).}
\end{figure}

For the experiments where $p=1$ and $n=200$, Figure~\ref{fig:boxplot_phi_n200_p1} provides a summary of results for the indicators $\vec{\theta}_{1,d}$ characterising the sparsity patterns in $\matr{\phi}_1$. Corresponding plots for the other five combinations of $n \in \{200, 1000\}$ and $p \in \{ 1, 2, 4\}$ are given in Supplementary Figure~S2. When stationarity is enforced through the prior, misclassification rates are consistently lower when there is a smaller degree of posterior uncertainty in the classification. In these cases, most misclassification rates are near zero. Notably, this is not true when stationarity is not enforced. Here, the misclassification rates of the naive Bayes classifier are often similar to those calculated using only the indicators whose value is more certain \textit{a posteriori}, especially for the higher orders, $p=2,4$. This deterioration in identification of sparsity when stationarity is not enforced is discussed further in Section~\ref{subsubsec:joint_posterior}.
%This suggests that when data arise from a stationary process but inference is not limited to the stationary region $\mathcal{C}_{p,m}$, the posterior can become unduly concentrated in regions of high likelihood, perhaps because most of the prior mass lies outside $\mathcal{C}_{p,m}$, leading to a smaller prior expectation of the likelihood.

Comparing now the Bayes classifiers for the two Bayesian analyses and the Lasso-based estimates, it is clear that the greatest difference in misclassification rates occurs when the proportion of sparsity is low. Here, misclassification rates are markedly lower in the Bayesian analyses when inference is limited to $\mathcal{C}_{p,m}$. When stationarity is not enforced, the Bayesian analyses and Lasso-based estimates produce similar results, with the former typically outperforming the latter when $p=1$ and conversely for $p=4$. As remarked in Section~\ref{sec:intro}, because of the high dimension of the parameter space, it can be difficult to learn that a process is stationary with certainty, especially when the sample size is small. This is borne out by the Bayesian analyses where stationarity is not enforced, with 10\% of the 540 analyses yielding a posterior for $\myPhi$ where up to 16\% of the mass lies outside $\mathcal{C}_{p,m}$. Typically, this occurred in experiments with a low proportion of zeros among the model parameters.

When the proportion of sparsity in the data-generating matrices is around 50\% or 90\%, the misclassification rates in the Bayesian analyses where stationarity is enforced are generally similar to those when stationarity is not enforced, with both typically outperforming the Lasso-based estimates. Perhaps unsurprisingly, it is noticeable that misclassification rates are generally much lower when there is a high proportion of sparsity in the model parameters.

Figure~\ref{fig:boxplot_K_n200_p1} shows a summary of the results concerning sparsity in the error precision matrix $\matr{K}$ for the experiments where $p=1$ and $n=200$. Corresponding plots for the other five combinations of $n \in \{200, 1000\}$ and $p \in \{ 1, 2, 4\}$ are provided in Supplementary Figure~S3. Across all simulation settings, misclassification rates based on the Bayesian analyses are very similar whether or not stationarity is enforced through the prior. In contrast, misclassification rates are markedly higher for the Lasso-based estimator. It is reassuring to observe that misclassification rates for both Bayesian analyses are lower when we only consider indicators whose values are more certain in the posterior.

\subsubsection{Joint posterior for the mixed graph $G$\label{subsubsec:joint_posterior}}

Supplementary Section~S5.2 presents an investigation into the joint posterior distribution for the mixed graph $G$ underpinning the model. A core result is that when the degree of sparsity in $\myPhi$ is small ($r=0.1$) or moderate ($r=0.5$), the posterior for the (Hamming) distance between $G_1=(V,E_1)$ and the subgraph used in simulating the data is typically markedly smaller when stationarity is enforced compared to when it is not. We argue that this may be due to prior-data conflict when stationarity is not enforced, arising from the induced prior for $\rho(\matr{C}_\phi)$ placing virtually all its mass where $\rho(\matr{C}_\phi)>1$; empirical evidence suggests that this may lead to compensatory inflation of the number of zeros in $\matr{C}_\phi$, thereby increasing the Hamming distance.

\section{\label{sec:applications}Applications}

\subsection{U.S. macroeconomic data}\label{subsec:macroeconomics}

In this section we consider an application involving a quarterly time series of U.S. macroeconomic variables. We focus primarily on demonstrating the improvements in predictive performance gained by using a prior that enforces stationarity and allows sparsity. 

The original dataset is described in \citet{Koo13}. After applying log-transformations and differencing, where appropriate, and then standardising all variables, the data are modelled as arising from a stationary, zero mean vector autoregression of order $p=4$. Following \citet{Koo13}, we construct a subset of $m=20$ variables whose names and mnemonics can be found in Supplementary Table~S7. Of these, interest lies in forecasting the first three  -- real gross domestic product (GDP251), the consumer price index (CPIAUCSL) and the Federal funds rate (FYFF) -- with the remaining seventeen variables included for their potential forecasting value. The data run from quarter 1 of 1959 to quarter 4 of 2007. In order to assess the forecasting properties of various prior-model combinations, we use the data for the first $n=156$ quarters for model-fitting, holding back the last 40 observations to allow assessment of out-of-sample predictive performance.

To this end, we consider a \VAR{20}{4} process under five prior distributions. Except in the first case, where the error precision matrix $\matr{K}$ is fixed, the autoregressive coefficients $\myPhi$ and $\matr{K}$ are taken to be independent \textit{a priori}:
\begin{enumerate}[(i)]
\item A Minnesota prior \citep[][]{DLS84} where the error variance matrix $\matr{\Sigma}$ is replaced with a fixed (diagonal) estimate and $\myPhi$ is given a multivariate normal distribution. Judicious choice of the prior mean and variance for $\myPhi$ is intended to encourage shrinkage towards a simple set of low order univariate AR models.
\item A semi-conjugate prior in which $\myPhi$ and $\matr{K}$ are given multivariate normal and Wishart distributions, respectively.
\item A prior for $\myPhi$ that enforces stationarity coupled with a Wishart prior for $\matr{K}$;  neither distribution allows sparsity in the parameter space. The prior for $\myPhi$ is induced through a prior on the partial autocorrelation matrices of the process \citep[][]{Hea23}.
\item A sparse prior that does not enforce stationarity. As in the simulation experiments, this comprises spike-and-slab priors for the off-diagonal elements of the $\matr{\phi}_s$ and a prior comprising a mixture of $G$-Wishart distributions for $\matr{K}$.
\item Our sparsity-inducing, stationarity-enforcing prior.
\end{enumerate}
Clearly, priors (i), (ii) and (iv) do not constrain inference to the stationary region, while priors (i)-(iii) do not allow sparsity. Only prior (v) removes both of these limitations. Full details of all prior specifications can be found in Supplementary Section~S6.1.2.

Priors (i)-(iii) were used in an analysis of the same data in \citet{Hea23}, where direct sampling from the analytically tractable posterior under prior (i) and MCMC sampling under non-conjugate priors (ii) and (iii) is described in full. Under priors (iv) and (v), we ran two chains, each generating 25K samples from the posterior, following a burn-in of 25K draws, and thinning to every tenth iterate to reduce post-processing overheads. The usual graphical diagnostics gave no evidence of any lack of convergence. It is worth noting that under the three priors which did not enforce stationarity, no posterior samples lay within the stationary region under the two non-sparse priors, (i) and (ii), while all the samples under the sparse prior (iv) lay within the stationary region. This demonstrates the important regularising effect that sparsity can have, ruling out implausible regions of parameter space when stationarity is a credible assumption.

Assessment of the predictive performance of the five prior-model combinations was based on forecast horizons of $h \in \{1,2,4,8\}$ time-steps, ranging from short-term, one-quarter ahead forecasts ($h=1$) to longer-term, two-year ahead forecasts ($h=8$). We use a number of proper scoring rules \citep[][]{GR07} to compare $h$-step ahead predictive distributions to the observations that materialised. At every $t=n+h, \ldots, n+40$ in the hold-out period the scores are based on the $h$-step ahead posterior predictive distributions at time $t$, which are then averaged. To assess forecasting performance for the three variables of interest individually we consider two widely-used scores: the continuous rank probability score (CRPS) and the logarithmic score. The energy score (ES), which is a multivariate generalisation of the CRPS, is then used to assess joint forecasts of the three variables of interest. All scores are negatively oriented, with small values indicating better performance. Further details on the scoring rules, along with additional comparisons, can be found in Supplementary Section~S6.1.3. 

\begin{figure}[t]
\centering
\includegraphics[width=0.8\textwidth]{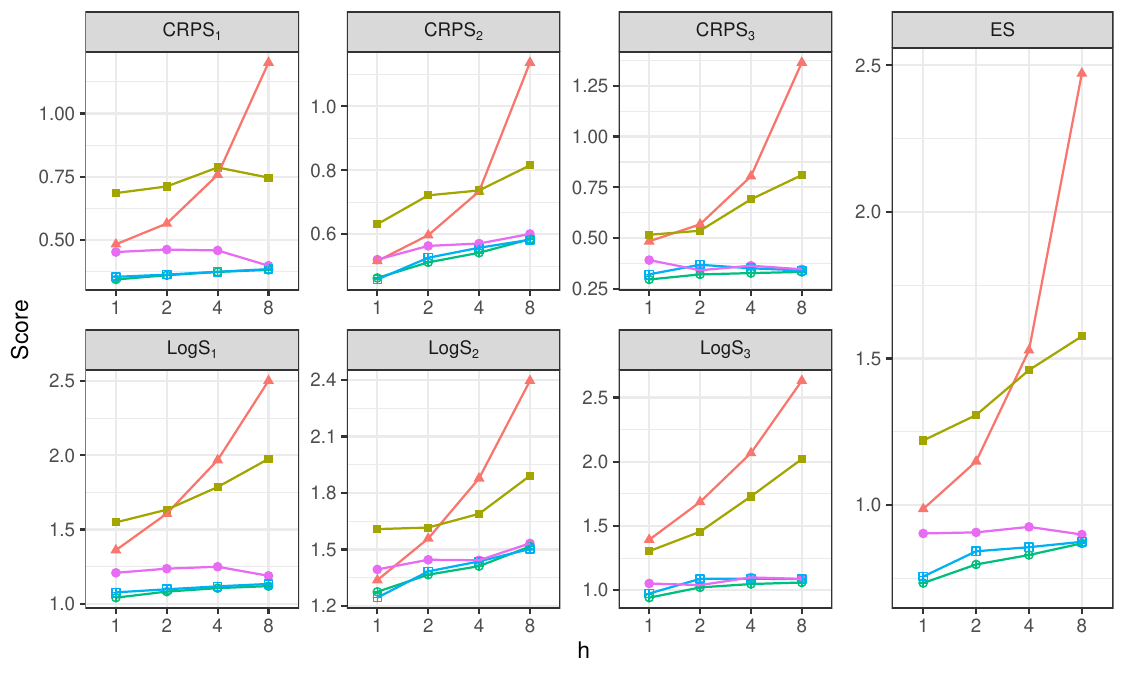}
\caption{\label{fig:macroforecasts_grid_nolegend}For each forecast horizon $h$ and each prior: $h$-step ahead CRPS and logarithmic score for variable $k$ (CRPS$_k$, LogS$_k$) and $h$-step ahead energy score (ES). Variables $k=1,2,3$ correspond to GDP251, CPIAUCSL and FYFF, respectively. The priors are represented through (i) \textcolor{ggplot5_1}{$\blacktriangle$} (ii) \textcolor{ggplot5_2}{$\blacksquare$} (iii) \textcolor{ggplot5_5}{$\bullet$} (iv) \textcolor{ggplot5_4}{$\boxplus$} (v) \textcolor{ggplot5_3}{$\oplus$}.}
\end{figure}

\begin{figure}[!t]
\centering
\includegraphics[width=0.75\textwidth, trim={0 3cm 0 3cm}, clip]{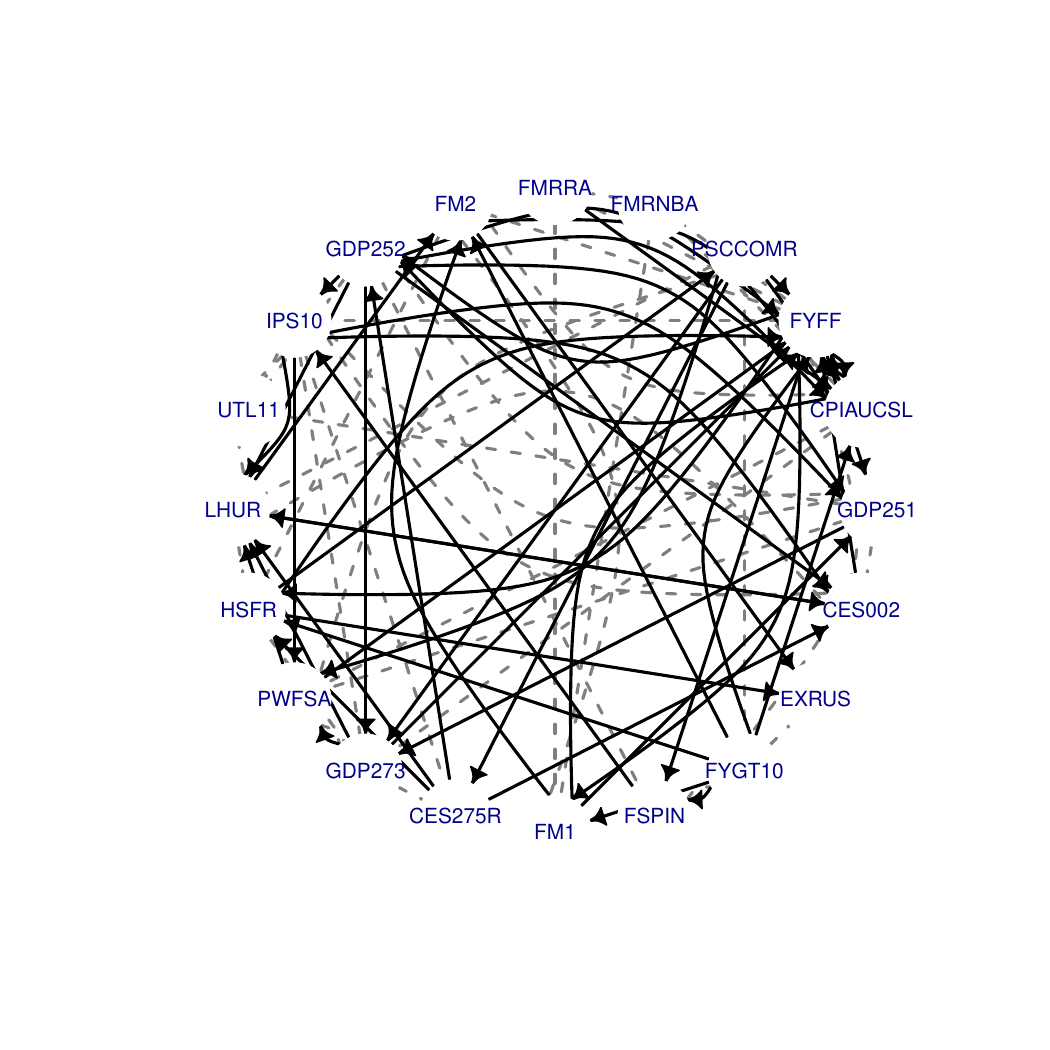}
\caption{\label{fig:macro_full_pathdiagram_joint}Mixed graph $G$ whose edges have posterior probability greater than 0.5 for the U.S. macroeconomic data. The graph shows directed ($\regulationarrow$) and undirected (\textcolor{gray}{\dashed}) edges.  The full variable names for the vertex labels are provided in Supplementary Table~S7.}
\end{figure}

Figure~\ref{fig:macroforecasts_grid_nolegend} shows the scores under each prior for each forecast horizon. It is immediately clear that the two non-sparse priors which do not constrain inference to the stationary region, (i) and (ii), produce the worst predictive performance, most notably at the larger horizons. This is because in both cases, the posterior probability that $\myPhi$ lies within the stationary region is approximately zero and so the variances of the predictive distributions grow without bound as the forecast horizon increases. Comparing the scores arising from the non-sparse priors (iii) and (ii) which do and do not enforce stationarity, respectively, the superior predictive performance in the former case is clear. An even more marked difference is seen comparing the scores under priors (iv) and (ii), neither of which enforce stationarity, but where sparsity is allowed only in the former case, yielding greatly improved predictive performance. However, for most scores and most forecast horizons, the best predictive performance arises under prior (v) which both restricts inference to the stationary region and allows sparsity in the parameter space. This is most noticeable in the individual scores for FYFF, which is the Federal funds rate. This is the variable which seems to interact most with the others. For example, in the posterior majority-rule sub-graph $G_1$, the sum of the in-degree and out-degree for FYFF is higher than that of all other nodes. This is illustrated in Figure~\ref{fig:macro_full_pathdiagram_joint} which shows the mixed graph, $G$, whose edges appeared with posterior probability greater than 0.5. Interestingly, among the most highly supported directed edges are those from CPIAUCSL (consumer price index) to FYFF, with posterior probability 0.9993, and from FYFF to CPIAUCSL, with posterior probability 0.9888. This suggests the existence of a feedback system between inflation and interest rates, in line with the results of \citet{ZW06}, Chapter 11, and \citet{Eic07}. After FYFF, the variable with the next highest sum of in-degrees and out-degrees is GDP252, which is real personal consumption. In terms of undirected edges, the variables with the highest degree are IPS10 (industrial production) and GDP251 (real gross domestic product). The large number of neighbours of FYFF and GDP252 in the directed graph $G_1$ and of IPS10 and GDP251 in the undirected graph $G_2$ are broadly consistent with the results from an analysis of the same data by \citet{ABC16}. More detailed information, in the form of posterior probabilities for all directed and undirected edges, are presented in Supplementary Figure~S6.

\subsection{Brain connectivity}\label{subsec:brain}

Understanding brain connectivity is a fundamental challenge in neuroscience. In this application, we apply our modelling procedure to a dataset of long-term intracranial electroencephalography (iEEG) recordings and consider implications of the implied sparsity structure. The data are described in detail in \citet{BHPWW24}. Briefly, the observations comprise pre-processed band power in two common frequency bands (beta and delta) from four individuals with refractory focal epilepsy, coded A, B, C and D. There are band power measurements for each of $m$ brain regions, averaged over 30 second segments, for $n$ time-steps during day-time hours. The dimension and length of the time-series varied between subjects with $m=8,8,9$ and $13$ and $n=685,622,651$ and $231$ for individuals A, B, C and D, respectively. The data were standardised prior to analysis and graphical interrogation suggested that stationarity was a plausible assumption. Subject C is unique in having data from bilateral sensors in both left and right hemispheres; we therefore focus much of our discussion on the results for this individual.

Previous analysis of these data \citep[][]{BHPWW24} suggested an appropriate choice of order $p$ for the stationary vector autoregressions; for instance $p=2$ in both the beta and delta bands for individual C. Moreover, preliminary analysis with similar data suggested a non-zero mean for the diagonal elements of $\matr{Z}_s$, $s=1,\ldots,p$, and so we use the version of the prior from Section~\ref{subsubsec:prior_zu} in which the prior mean $\mu$ of the diagonal effect sizes $\tilde{z}_{s,ii}$ is unknown and assigned a normal prior, $\mu \sim \norm(f_1, f_2^2)$. A complete specification of the prior hyperparameters can be found in Supplementary Section~S6.2.1.

For each dataset, we ran two chains of the MCMC sampler described in Section~\ref{sec:mcmc}, each generating 250K samples from the posterior, following a burn-in of 250K draws, and thinning to every hundredth iterate to reduce post-processing overheads. The usual graphical diagnostics gave no evidence of any lack of convergence.

\begin{figure}[t]
\centering
\includegraphics[width=0.85\textwidth]{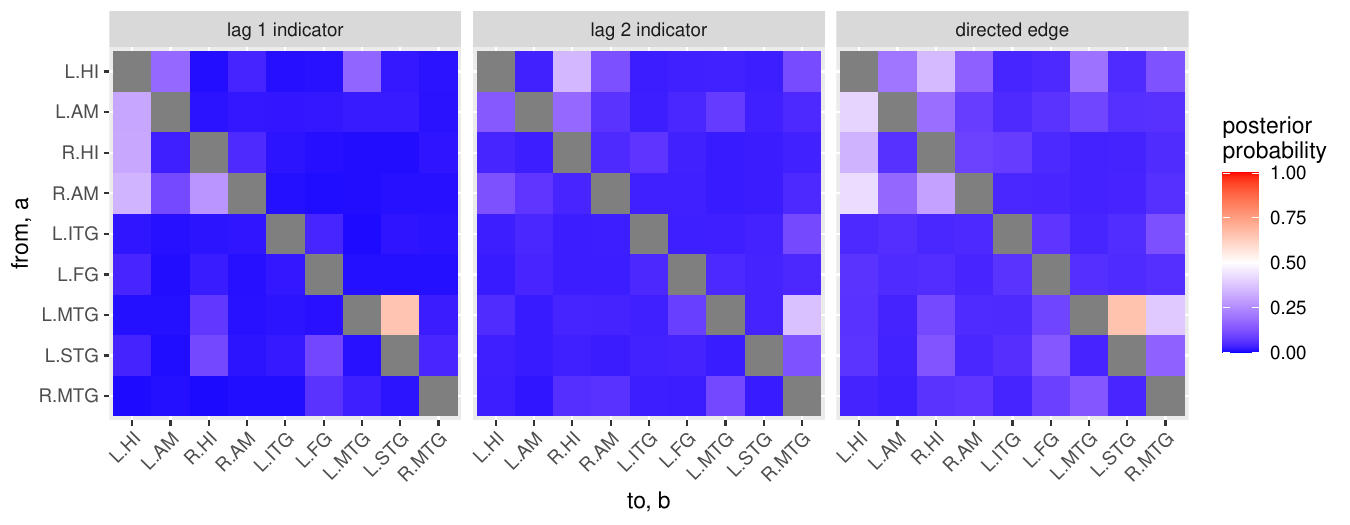}
\caption{\label{fig:beta_909_flat_diredge_probs_joint}For data in the beta band for individual C, posterior probabilities of the off-diagonal indicators in $\matr{\Gamma}_1^\T=(\gamma_{1,ba})^\T$ (left) and $\matr{\Gamma}_2^\T=(\gamma_{2,ba})^\T$ (middle) and the directed edges $a \diredge b$ (right). Vertex labels are prefixed by L. or R. to indicate whether the brain region is in the left or right hemisphere.}
\end{figure}

\begin{figure}[t]
\centering
\includegraphics[width=0.7\textwidth]{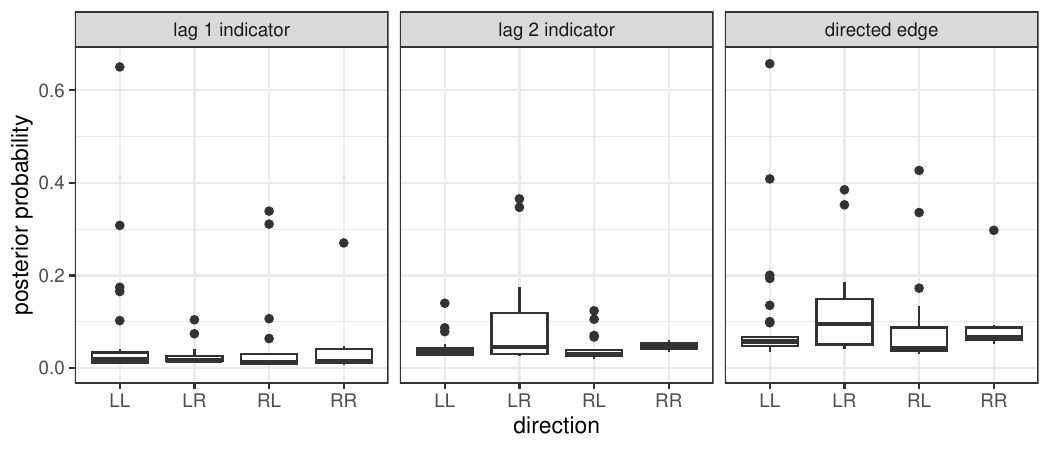}
\caption{\label{fig:brain_beta_909_flat_prob_vs_direction}For data in the beta band for individual C, boxplots summarising posterior probabilities of the off-diagonal indicators in $\matr{\Gamma}_1=(\gamma_{1,ba})$ (left) and $\matr{\Gamma}_2=(\gamma_{2,ba})$ (middle) and the directed edges $a \diredge b$ (right) according to whether $a \diredge b$ corresponds to the LL, LR, RL or RR direction.}
\end{figure}

A plot showing the locations of the brain regions for individual C is provided in Supplementary Figure~S7; there are six regions in the left hemisphere and three in the right. For the band power data in the beta band, Figure~\ref{fig:beta_909_flat_diredge_probs_joint} visualises the posterior probabilities of the indicators $\vec{\theta}_{1,d}$ defining the patterns of sparsity in the autoregressive coefficient matrices along with the posterior probabilities of all possible directed edges. Although there are few directed edges with posterior probability greater than 0.5, evidence of structure becomes apparent when the indicators $\gamma_{s,ba}$, $s=1,2$, or edges $a \diredge b$, are grouped according to whether the regions corresponding to variables $a$ and $b$ are in the left or right hemispheres; we refer to these as left-to-left (LL), left-to-right (LR), right-to-left (RL) or right-to-right (RR) relationships or edges. To this end, Figure~\ref{fig:brain_beta_909_flat_prob_vs_direction} shows boxplots of the posterior probabilities associated with each category. It is clear from the right panel that the LR edges have markedly higher posterior probability than directed edges of the other three types. From the left and central panels, it seems this is due to markedly higher posterior probabilities of non-zero coefficients at lag 2, which corresponds to a lag of one minute. Corresponding plots for the data in the delta band are shown in Supplementary Figures~S8 and S9. The delta band showed no notable differences between the probabilities of LL, LR, RL and RR relationships or edges, possibly due to the relative prominence of these frequency bands in these brain regions \citep[][]{Tay20}. Nevertheless, the evidence of higher probabilities for LR edges in the beta band is interesting in the context of the epilepsy: this individual was shown to have epileptic seizures originating from the left hemisphere (amygdala and hippocampus), and after removal of the left temporal lobe during epilepsy surgery, this individual was subsequently seizure free. Interestingly, the band power data used for the analysis here did not contain seizures, and it is therefore a first indication that long-term (ultradian) brain activity during interictal (non-seizure) periods may also contain information about the seizure-generating functional network.
%Is there anything worth saying about the relationships seeming to be at lag 2 (lag of one minute) rather than lag 1 (lag of 30 seconds)? - not sure on this one! I think we'd need a more systematic look with different sampling frequencies to say?

For all individuals, the distances between the $m$ brain regions can be represented using Euclidean distance between the centroids of the volumetric brain regions. The brain regions are obtained as part of the standard Desikan-Killiany atlas from Freesurfer \citep[][]{Ale19}, and we used the standardised atlas space to obtain the Euclidean distances. We denote the distance between regions $a$ and $b$ by $D_{ab}$. For each of the four individuals and both bands, Table~\ref{tab:corr_dist_prob} shows the correlation  between $D_{ab}$ and the (logit transformed) posterior probabilities, $\Pr(a \diredge b | \vec{y})$ and $\Pr(a \edge b | \vec{y})$. From these statistics, it appears that there is a modest, negative association between the distance between brain regions and the posterior probability of an undirected edge between those regions, but little evidence of a linear association between distance and the posterior probability of a directed edge. This suggests that (i) spatial factors can partly explain the contemporaneous correlation between band power in different brain regions, but (ii) not the directed (lagged) relationships. The former is in keeping with previous literature on so called ``functional connectivity'' in these types of recordings \citep[][]{Wan20}. The primary reason for the spatial dependency structure of contemporaneous recordings is understood to be due to the volume conduction of electric fields, such that two nearby recording sites would ``see'' similar electric signals, and we suggest this is also the case here. The latter observation, instead, indicates that the directed (lagged) relationships uncovered here may represent a causal biological process, possibly related to the generation of epilepsy from pathological tissues.

\begin{table}
\caption{Pearson correlation coefficients between the distance $D_{ab}$ and $\mathrm{logit}\{\Pr(a \diredge b | \vec{y}\}$ for directed edges, and between $D_{ab}$ and $\mathrm{logit}\{\Pr(a \edge b | \vec{y})\}$ for undirected edges.} \label{tab:corr_dist_prob}\tabularnewline
\centering
\begin{tabular}{@{}ccccc@{}}
\toprule\noalign{}
Individual &\multicolumn{2}{c}{Beta band} &\multicolumn{2}{c}{Delta band}\\
\cmidrule(lr){2-3} \cmidrule(lr){4-5}
 &Directed &Undirected &Directed &Undirected\\
\midrule\noalign{}
A &-0.076 &-0.477 &-0.296 &-0.385\\
B &-0.001 &-0.089 &-0.077 &-0.214\\
C &-0.144 &-0.375 &0.046  &-0.257\\
D &-0.150 &-0.365 &-0.059 &-0.377\\
%A &beta  &-0.076 ($p=$0.526) &-0.477 ($p=$0.003)\\
%A &delta &-0.296 ($p=$0.012) &-0.385 ($p=$0.020)\\
%B &beta  &-0.001 ($p=$0.993) &-0.089 ($p=$0.651)\\
%B &delta &-0.077 ($p=$0.571) &-0.214 ($p=$0.274)\\
%C &beta  &-0.144 ($p=$0.289) &-0.375 ($p=$0.049)\\
%C &delta &0.046  ($p=$0.736) &-0.257 ($p=$0.186)\\
%D &beta  &-0.150 ($p=$0.061) &-0.365 ($p=$0.001)\\
%D &delta &-0.059 ($p=$0.466) &-0.377 ($p=$0.001)\\
\bottomrule\noalign{}
%\end{longtable}
\end{tabular}
\end{table}

\section{Discussion}\label{sec:discussion}

For many applications, it can be advantageous to assume that a time-series is stationary, for example when structural or functional constraints on the process mean the predictive variance should be bounded. In analyses based on graphical vector autoregressions, stationarity is frequently stated as a model assumption but rarely enforced as a constraint. In this paper we provide, to the best of our knowledge, the first methodology for constructing a prior for the autoregressive coefficients of a vector autoregression which simultaneously constrains inference to the stationary region and allows sparsity in the parameter space. This is achieved by introducing an auxiliary variable and then transforming the expanded set of autoregressive coefficient matrices to an expanded set of unconstrained square matrices through a mapping which preserves zeros. By assigning a sparse prior to the unconstrained square matrices, a sparse prior is induced for the autoregressive coefficient matrices that guarantees the process is stationary. Completing the prior specification with a mixture of $G$-Wishart distributions for the error precision matrix yields a prior for the parameters of a stationary graphical vector autoregression. 

An MCMC scheme for computational inference is presented which combines proposals using NUTS for continuous model parameters with an efficient reversible jump sampler for the error precision matrix and its sparsity indicators. A link to software implementing this algorithm is provided in the Supplementary Material.

Through a comprehensive simulation experiment we demonstrate that when stationarity is a reasonable assumption, characterisation of posterior uncertainty in sparsity patterns is generally improved when a sparse prior that restricts inference to the stationary region is used. We posit that this may be due to improved prior-data agreement, which limits confounding between the spectral radius of the companion matrix $\matr{C}_\phi$ and the number of zeros it contains. In an application involving U.S. macroeconomic data, we demonstrate the improvements in predictive performance that can be gained by adopting a prior that limits inference to the stationary region, allows sparsity in the parameter space, or both, with the greatest improvement in predictive performance achieved in the third case. Finally, an application involving brain activity data demonstrates the scientific insights that can be gained from the posterior for the directional indicators defining the patterns of sparsity in our stationary graphical vector autoregression.

We emphasise that the ideas in this paper could be adapted readily  to develop sparsity-inducing priors over other spaces subject to norm-like constraints. This includes the simplex, with implications for mixture modelling and model-based clustering.

\section*{Acknowledgements}
All authors were supported by the EPSRC grant UKRI2402. This work has made use of the Hamilton HPC Service of Durham University.

%\section{Data Availability Statement}\label{data-availability-statement}

%Deidentified data have been made available at the following URL: XX.

\section*{Supplementary Materials}

\begin{description}
\item[Supplement to ``Bayesian inference of sparsity in stable vector autoregressive processes'':]
File of supplementary information including derivation of the Jacobian determinant for the mapping from the constrained to the unconstrained parameter spaces, further theory on the shrinkage properties of the prior, additional details of the MCMC scheme for sampling from the posterior distribution and extra details, figures and tables for the simulation experiment and applications. (.pdf file).
\item[Python code to fit the stationary graphical vector autoregression:]
The simulated data from Section~\ref{sec:sim_expt} and Python scripts for fitting the model can be found at the GitHub repository \url{https://github.com/nseg4/bayes-sparse-stable-var}.
\end{description}

\bibliography{journal_names_abbr.bib,refs.bib}

@STRING(BIOK = "Biometrika")

@STRING(ECON = "Econometrica")

@STRING(JASA = "J. Amer. Statist. Assoc.")

@STRING(JSCS = "J. Statist. Comput. Simul.")

@STRING(JE = "J. Economet.")

@STRING(JSS = "J. Statist. Software")

@STRING(CSDA = "Comput. Statist. Data Anal.")

@STRING(JAE = "J. Appl. Economet.")

@STRING(JTSA = "J. Time Ser. Anal.")

@STRING(BA = "Bayesian Anal.")

@STRING(HBM = "Hum. Brain Mapp.")

@STRING(FCN = "Front. Comput. Neurosc.")

@STRING(ER = "Economet. Rev.")

@STRING(JMLR = "J. Mach. Learn. Res.")

@STRING(ML = "Mach. Learn.")

@STRING(JCGS = "J. Comput. Graph. Stat.")

@STRING(JBES = "J. Bus. Econ. Stat.")

@ARTICLE{Hea23,
  author = "S.~E.~Heaps",
  title = "Enforcing Stationarity through the Prior in Vector Autoregressions",
  journal = JCGS,
  year = "2023",
  volume = "32",
  number = "1",
  pages = "74--83"
}

@ARTICLE{Eic12,
  author = "M.~Eichler",
  title = "Graphical modelling of multivariate time series",
  volume = "153",
  journal = {Probability Theory and Related Fields},
  pages = "233--268",
  year = "2012"
}

@ARTICLE{Gra69,
  author = "C.~W.~J.~Granger",
  title = "Investigating causal relations by econometric models and cross-spectral methods",
  volume = "37",
  journal = ECON,
  pages = "24--36",
  year = "1969"
}

@ARTICLE{CGYHSV17,
  author = "S.~Chiang and M.~Guindani and H.~J.~Yeh and Z.~Haneef and J.~M.~Stern and M.~Vannucci",
  title = "Bayesian vector autoregressive model for multi-subject effective connectivity inference using multi-modal neuroimaging data",
  journal = HBM,
  pages = "1311--1332",
  year = "2017",
  volume = "38"
}

@ARTICLE{GFOBC13,
  author = "C.~Gorrostieta and M.~Fiecas and H.~Ombao and E.~Burke and S.~Cramer",
  title = "Hierarchical vector auto-regressive models and their applications to multi-subject effective connectivity",
  journal = FCN,
  pages = "159",
  year = "2013",
  volume = "7"
}

@ARTICLE{PC20,
  author = "L.~Paci and G.~Consonni",
  title = "Structural learning contemporaneous dependencies in graphical {VAR} models",
  journal = CSDA,
  pages = "106880",
  year = "2020",
  volume = "144"
}

@ARTICLE{CV05,
  author = "J.~Corander and M.~Villani",
  title = "A {B}ayesian approach to modelling graphical vector autoregressions",
  journal = JTSA,
  pages = "141--156",
  year = "2005",
  volume = "27",
  number = "1"
}

@ARTICLE{HSW13,
  author = "Y.~He and Y.~She and D.~Wu",
  title = "Stationary-sparse causality network learning",
  journal = JMLR,
  volume = "14",
  number = "58",
  pages = "3073--3104",
  year = "2013"
}

@ARTICLE{SF22,
  author = "A.~Shojaie and E.~B.~Fox",
  title = "Granger causality: a review and recent advances",
  journal = {Annual Review of Statistics and its Application},
  volume = "9",
  pages = "289--319",
  year = "2022"
}

@ARTICLE{DYMZ23,
  author = "L.~L.~Duan and Z.~Yuwen and G.~Michailidis and Z.~Zhang",
  title = "Low tree-rank {B}ayesian vector autoregression model",
  journal = JMLR,
  volume = "24",
  pages = "1--35",
  year = "2023"
}

@ARTICLE{FSCS22,
  author = "J.~Fan and K.~Sitek and B.~Chandrasekaran and A.~Sarkar",
  title = "Bayesian tensor factorized vector autoregressive models for inferring {G}ranger causality patterns from high-dimensional multi-subject panel neuroimaging data",
  journal = {arXiv:2206.10757},
  year = "2022",
  adsurl = {https://arxiv.org/abs/2206.10757},
}

@INCOLLECTION{MO20,
  author    = "S.~Mukherjee and C.~Oates",
  title     = "Graphical models in molecular systems biology",
  booktitle = "Handbook of {G}raphical {M}odels",
  publisher = {Chapman \& Hall/CRC},
  year      = "2020",
  editor    = "M.~Maathuis and M.~Drton and S.~Lauritzen and M.~Wainwright",
  pages     = "497--511"
}

@ARTICLE{MA13,
  author = "G.~Michailidis and F.~{d'Alch\'{e}-Buc}",
  title = "Autoregressive models for gene regulatory network inference: sparsity, stability and causality issues",
  volume = "246",
  journal = {Mathematical Biosciences},
  pages = "326--334",
  year = "2013"
}

@ARTICLE{ABC16,
  author = "D.~F.~Ahelegbey and M.~Billio and R.~Casarin",
  title = "Bayesian graphical models for structural vector autoregressive processes",
  journal = {Journal of Applied Econometrics},
  volume = "31",
  pages = "357--386",
  year = "2016"
}

@ARTICLE{Rov02,
  author = "A.~Roverato",
  title = "Hyper inverse {W}ishart distribution for non-decomposable graphs and its application to {B}ayesian inference for {G}aussian graphical models",
  volume = "29",
  journal = {Scandinavian Journal of Statistics},
  pages = "391--411",
  year = "2002"
}

@ARTICLE{AM05,
  author = "A.~{Atay-Kayis} and H.~Massam",
  title = "A {M}onte {C}arlo method for computing the marginal likelihood in nondecomposable {G}aussian graphical models",
  volume = "92",
  number = "2",
  journal = BIOK,
  pages = "317--335",
  year = "2005"
}

@ARTICLE{HG14,
  author = "M.~D.~Hoffman and A.~Gelman",
  title = "The {N}o-{U}-{T}urn {S}ampler: adaptively setting path lengths in {H}amiltonian {M}onte {C}arlo",
  year = "2014",
  journal = JMLR,
  volume = "15",
  pages = "1593--1623"
}

@ARTICLE{GSN08,
  author = "E.~I.~George and D.~Sun and S.~Ni",
  title = "Bayesian stochastic search for {VAR} model restrictions",
  journal = JE,
  volume = "142",
  pages = "553--580",
  year = "2008"
}

@ARTICLE{LBGGW11,
  title = "Bayesian inference for sparse {VAR}(1) models, with application to time course microarray data",
  author = "G.~Lei and R.~J.~Boys and C.~S.~Gillespie and A.~J.~Greenall and D.~J.~Wilkinson",
  journal = {Journal of Biometrics and Biostatistics},
  year = "2011",
  volume = "2",
  number = "127"
}

@ARTICLE{HHNCAGW23,
  title = "A sparse {B}ayesian hierarchical vector autoregressive model for microbial dynamics in a wastewater treatment plant",
  author = "N.~E. Hannaford and S.~E.~Heaps and T.~M.~W.~Nye and T.~P. Curtis and B.~Allen and A.~Golightly and D.~J.~Wilkinson",
  journal = CSDA,
  year = "2023",
  volume = "179",
  number = "127",
  pages = "107659"
}

@ARTICLE{RRG24,
  author = "A.~Roy and A.~Roy and S.~Ghosal",
  title = "Bayesian inference for relational graph in a causal vector autoregressive time series",
  journal = {arXiv:2410.22617},
  year = "2024",
  adsurl = {https://arxiv.org/abs/2410.22617}
}

@ARTICLE{MC09,
  title = "Bayesian learning of graphical vector autoregressions with unequal lag-lengths",
  author = "P. Marttinen and J.~Corander",
  journal = ML,
  year = "2009",
  volume = "75",
  pages = "217--243"
}

@ARTICLE{BBB24,
  title = "Variational inference for large {B}ayesian vector autoregressions",
  author = "M.~Bernardi and D.~Bianchi and N.~Bianco",
  journal = JBES,
  year = "2024",
  volume = "42",
  number = "3",
  pages = "1066--1082"
}

@ARTICLE{Kor16,
  title = "Prior selection for panel vector autoregressions",
  author = "D.~Korobilis",
  journal = CSDA,
  year = "2016",
  volume = "101",
  pages = "110--120"
}

@ARTICLE{MV99,
   author = "{X.-L.}~Meng and D.~A.~{Van Dyke}",
   title = "Seeking efficient data augmentation schemes via conditional and marginal augmentation",
   journal = BIOK,
   volume = "86",
   number = "2",
   pages = "301--320",
   year  = "1999"
}

@ARTICLE{LW99,
   author = "J.~S.~Liu and Y.~N.~Wu",
   title = "Parameter expansion for data augmentation",
   journal = JASA,
   volume = "94",
   number = "448",
   pages = "1264--1274",
   year  = "1999"
}

@ARTICLE{JHD21,
  author = "M.~Jauch and P.~D.~Hoff and D.~B.~Dunson",
  title = "Monte {C}arlo simulation on the {S}tiefel manifold via polar expansion",
  journal = JCGS,
  volume = "30",
  number = "3",
  pages = "622--631",
  year = "2021"
}

@ARTICLE{HJ24,
  author = "S.~E.~Heaps and I.~H.~Jermyn",
  title = "Structured prior distributions for the covariance matrix in latent factor models",
  journal = {Statistics and Computing},
  volume = "34",
  number = "143",
  year = "2024"
}

@MISC{JAX18,
  author = "J.~Bradbury and R.~Frostig and P.~Hawkins and M.~J.~Johnson and C.~Leary and D.~Maclaurin and G.~Necula and A.~Paszke and J.~Vander{P}las and S.~Wanderman-{M}ilne and Q.~Zhang",
  title = "{JAX}: composable transformations of {P}ython+{N}um{P}y programs",
  note = "Version 0.3.13",
  year = "2018",
}

@ARTICLE{BLACKJAX24,
  title = "Black{JAX}: Composable {B}ayesian inference in {JAX}",
  author = "A.~Cabezas and A.~Corenflos and J.~Lao and R.~Louf",
  year = "2024",
  journal = {arXiv:2402.10797}
}

@INCOLLECTION{Nea11,
  author = "R.~M.~Neal",
  booktitle = {Handbook of Markov Chain Monte Carlo},
  editor = "S.~Brooks and A.~Gelman and G.~Jones and X.--L.~Meng",
  pages = "113--162",
  title = "{MCMC} Using {H}amiltonian Dynamics",
  year = "2011",
  publisher = {Chapman \& Hall/CRC},
  series = "Handbooks of modern statistical methods"
}

@INCOLLECTION{MGM06,
  author = "I.~Murray and Z.~Ghahramani and D.~J.~C.~MacKay",
  title = "{MCMC} for doubly-intractable distributions",
  booktitle = {Proceedings of the 22nd Annual Conference on Uncertainty in Artificial Intelligence (UAI-06)},
  publisher = {AUAI Press},
  year = "2006",
  pages = "359--366"
}

@ARTICLE{CGH17,
  title = "Stan: A probabilistic programming language",
  author = "B.~Carpenter and A.~Gelman and M.~D.~Hoffman and D.~Lee and B.~Goodrich and M.~Betancourt and M.~A.~Brubaker and J.~Guo and P.~Li and A.~Riddell",
  journal = JSS,
  year = "2017",
  volume = "76",
  number = "1",
  pages = "1--32"
}

@ARTICLE{VMS24,
  author = "L.~Vogels and R.~Mohammadi and M.~Schoonhoven and {\c{S}}.~{\.{I}}.~Birbil",
  title = "Bayesian Structure Learning in Undirected {G}aussian Graphical Models: Literature Review with Empirical Comparison",
  journal = JASA,
  volume = "119",
  number = "548",
  pages = "3164--3182",
  year = "2024"
}

@INCOLLECTION{Mas20,
  author    = "H.~Massam",
  title     = "Bayesian inference in graphical {G}aussian models",
  booktitle = "Handbook of {G}raphical {M}odels",
  publisher = {Chapman \& Hall/CRC},
  year      = "2020",
  editor    = "M.~Maathuis and M.~Drton and S.~Lauritzen and M.~Wainwright",
  pages     = "239--263"
}

@ARTICLE{WL12,
  author = "H.~Wang and S.~Z.~Li",
  title = "Efficient {G}aussian graphical model determination under {$G$}-{W}ishart prior distributions",
  journal = {Electronic Journal of Statistics},
  volume = "6",
  pages = "168--198",
  year = "2012"
}

@ARTICLE{CC08,
  author = "J.~Chen and Z.~Chen",
  title = "Extended {B}ayesian information criteria for model selection with large model spaces",
  journal = BIOK,
  volume = "95",
  number = "3",
  pages = "759--771",
  year = "2008"
}

@MANUAL{Eps24,
  title = "graphical{VAR}: graphical {VAR} for experience sampling data",
  author = "S.~Epskamp",
  year = "2024",
  note = {R package version 0.3.4}
}

@ARTICLE{Koo13,
  author = "G.~M.~Koop",
  title = "Forecasting with Medium and Large {B}ayesian {VAR}s",
  journal = JAE,
  volume = "28",
  number = "2",
  pages = "177--203",
  year = "2013"
}

@ARTICLE{BHPWW24,
  author = "R.~L.~Binks and S.~E.~Heaps and M.~Panagiotopoulou and Y.~Wang and D.~J.~Wilkinson",
  title = "Bayesian inference on the order of stationary vector autoregressions",
  journal = BA,
  pages = "1--22",
  year = "2024"
}

@ARTICLE{GR07,
  title = "Strictly proper scoring rules, prediction, and estimation",
  author = "T.~Gneiting and A.~E.~Raftery",
  journal = JASA,
  year = "2007",
  volume = "102",
  number = "477",
  pages = "359--378"
}

@ARTICLE{DLS84,
  author = "T.~Doan and R.~B.~Litterman and C.~A.~Sims",
  title = "Forecasting and conditional projection using realistic prior distributions",
  journal = ER,
  volume = "3",
  number = "1",
  pages = "1--100",
  year = "1984"
}

@ARTICLE{Eic07,
  title = "Granger causality and path diagrams for multivariate time series",
  author = "M.~Eichler",
  journal = JE, 
  year = "2007",
  volume = "137",
  pages = "334--353"
}

@BOOK{ZW06,
  title = "Modelling Financial Time Series with {S-PLUS}",
  author = "E.~Zivot and J.~Wang",
  publisher = "Springer",
  year = "2006"
}

@ARTICLE{Lia10,
  title = "A double {M}etropolis-{H}astings sampler for spatial models with intractable normalizing constants",
  author = "F.~Liang",
  journal = JSCS, 
  year = "2010",
  volume = "80",
  number = "9",
  pages = "1007--1022"
}

@PHDTHESIS{Luo25,
  author  = "Y.~Luo",
  title   = "{P}arsimonious {T}ime {S}eries {M}odelling of {H}igh-dimensional {D}ata with {L}inear and {N}on-{L}inear {M}odels",
  school  = "University College London",
  year    = "2025"
}

@ARTICLE{Tay20,
  author = "P.~N.~Taylor and C.~A.~Papasavvas and T.~W.~Owen and G.~M.~Schroeder and F.~E.~Hutchings and F.~A.~Chowdhury and B.~Diehl and J.~S.~Duncan and A.~W.~Mc{E}voy and A.~Miserocchi and J.~{de Tisi} and S.~B.~Vos and M.~C.~Walker and Y.~Wang",
  title = "Normative brain mapping of interictal intracranial {EEG} to localize epileptogenic tissue",
  volume = "145",
  journal = {Brain},
  number = "3",
  pages = "939--949",
  year = "2020"
}

@ARTICLE{Ale19,
  author = "B.~Alexander and W.~Y.~Loh and L.~G.~Matthews and A.~L.~Murray and C.~Adamson and R.~Beare and J.~Chen and C.~E.~Kelly and P.~J.~Anderson and L.~W.~Doyle and A.~J.~Spittle and J.~L.~Y.~Cheong and M.~L.~Seal and D.~K.~Thompson",
  title = "Desikan-{K}illiany-{T}ourville atlas compatible version of {M-CRIB} neonatal parcellated whole brain atlas: the {M-CRIB} 2.0",
  volume = "13",
  journal = {Frontiers in Neuroscience},
  pages = "34",
  year = "2019"
}

@ARTICLE{Wan20,
  author = "Y.~Wang and N.~Sinha and G.~M.~Schroeder and S.~Ramaraju and A.~W.~Mc{E}voy and A.~Miserocchi and J.~{de Tisi} and F.~A.~Chowdhury and B.~Diehl and J.~S.~Duncan and P.~N.~Taylor",
  title = "Interictal intracranial electroencephalography for predicting surgical success: the importance of space and time",
  volume = "61",
  journal = {Epilepsia},
  number = "7",
  pages = "1417--1426",
  year = "2020"
}

\end{document}